\documentclass[nature,superscriptaddress,showpacs]{revtex4-1}
\usepackage{textcomp}
\usepackage{latexsym}
\usepackage{amsthm}
\usepackage{amssymb}
\usepackage{amsmath}
\usepackage[all]{xy}
\usepackage{graphicx}
\usepackage{dcolumn}
\usepackage{bm}
\usepackage{graphicx}
\usepackage{epstopdf}

\usepackage{multirow}
\usepackage{color}
\definecolor{Gray}{gray}{0.9}
\definecolor{LightCyan}{rgb}{0.88,1,1}
\usepackage[first=0,last=9]{lcg}

\usepackage{setspace}

\SelectTips{eu}{11}

\begin{document}
\begin{spacing}{1.5}

\title{Experimental realization of quantum algorithm for solving linear systems of equations }

\author{Jian Pan}
\affiliation{Hefei National Laboratory for Physical Sciences at Microscale and Department of Modern Physics, University of Science and Technology of China, Hefei, 230026, People's Republic of China}

\author{Yudong Cao}
\affiliation{Department of Mechanical Engineering, Purdue University}

\author{Xiwei Yao}
\affiliation{Department of Electronic Science and Fujian Key Laboratory of Plasma and Magnetic Resonance, School of Physics and Mechanical and Electrical Engineering,Xiamen University, Xiamen, Fujian 361005, China.}

\author{Zhaokai Li}
\affiliation{Hefei National Laboratory for Physical Sciences at Microscale and Department of Modern Physics, University of Science and Technology
of China, Hefei, 230026, People's Republic of China}

\author{Chenyong Ju}
\affiliation{Hefei National Laboratory for Physical Sciences at Microscale and Department of Modern Physics, University of Science and Technology
of China, Hefei, 230026, People's Republic of China}

\author{Xinhua Peng}
%\email[E-mail me at: ]{xhpeng@ustc.edu.cn}
\affiliation{Hefei National Laboratory for Physical Sciences at Microscale and Department of Modern Physics, University of Science and Technology
of China, Hefei, 230026, People's Republic of China}

\author{Sabre Kais}
\email[E-mail me at: ]{kais@purdue.edu}
\affiliation{Department of Chemistry, Physics and Birck Nanotechnology Center, Purdue University,
West Lafayette, IN 47907, USA}

\author{Jiangfeng Du}
\email[E-mail me at: ]{djf@ustc.edu.cn}
\affiliation{Hefei National Laboratory for Physical Sciences at Microscale and Department of Modern Physics, University of Science and Technology
of China, Hefei, 230026, People's Republic of China}

\begin{abstract}
Quantum computers have the potential of solving certain problems exponentially faster than classical computers. Recently, Harrow, Hassidim and Lloyd proposed a quantum algorithm for solving linear systems of equations: given an $N\times{N}$ matrix $A$ and a vector $\vec b$, find the vector $\vec x$ that satisfies $A\vec x  = \vec b$. It has been shown that using the algorithm one could obtain the solution encoded in a quantum state $|x\rangle$ using $O(\log{N})$ quantum operations, while classical algorithms require at least $O(N)$ steps. If one is not interested in the solution $\vec{x}$ itself but certain statistical feature of the solution $\langle{x}|M|x\rangle$ ($M$ is some quantum mechanical operator), the quantum algorithm will be able to achieve exponential speedup over the best classical algorithm as $N$ grows. Here we report a proof-of-concept experimental demonstration of the quantum algorithm using a 4-qubit nuclear magnetic resonance (NMR) quantum information processor. For all the three sets of experiments with different choices of $\vec b$, we obtain the solutions with over $96\%$ fidelity. This experiment is a first implementation of the algorithm. Because solving linear systems is a common problem in nearly all fields of science and engineering, we will also discuss the implication of our results on the potential of using quantum computers for solving practical linear systems.
\end{abstract}


\maketitle

%\section{Introduction}\label{sec:intro}
The proposition of quantum computer dates back to 1980s \cite{feynman86}, but it was not until the late 80's and early 90's that quantum computers are shown to be more powerful than classical computers on various specialized problems \cite{DJ92,Grover96,Lloyd96,Shor94}. For example, the Deutsch-Jozsa algorithm \cite{DJ92}, Shor's quantum algorithm for factoring integers \cite{Shor94}, Grover's quantum search algorithm \cite{Grover96} and algorithms for Hamiltonian simulation of quantum systems \cite{Lloyd96} have been found to require significantly less computational steps than their classical counterparts and thus render many classically intractable problems realistically solvable with a quantum computer. There has been experimental demonstrations of important quantum algorithms such as Shor's algorithm~\cite{shor}, Grover search algorithm~\cite{grover}, optimization problems~\cite{chuang_opt}, quantum simulation of molecular systems~\cite{white_quan_chem,walther_quan_chem,baugh_quan_sim,cory_quan_sim}, all using small-scale quantum information processors. In this work we implement an algorithm for solving linear systems \cite{originalpaper} using NMR.

Linear systems of equations play an important role in nearly all fields of science and engineering. In quantum reactive scattering, the Kohn variational calculation involves the inversion of the augmented stiffness matrix \cite{levine_scattering,wyatt_scattering}, which is equivalent to solving a linear system in certain occasions. In chemistry, linear equations arise commonly in problems such as electrostatic calculation in density functional theory, where the discretized Poisson equation takes a linear form~\cite{parr_DFT}. Recently the Finite Element method starts to be adopted for solving electronic structure problems in quantum chemistry \cite{martinez_FEA}, where Schr\"{o}dinger's equation is recast in form of a linear equation. Also solving linear systems of equations often play a role as intermediate step in many algorithms such as quantum algorithm for data fitting \cite{datafitting}.

Here we consider the particular type of linear system where we are given an ${ N} \times { N}$ $s$-sparse Hermitian matrix $A$ with condition number $\kappa$ and unit vector $\vec{b}$ and we are interested not in the solution $\vec{x}$ itself but certain feature of $\vec{x}$ that can be written in form of $\vec{x}^\dagger{M}\vec{x}$ ($M$ is some linear quantum mechanical operator). As the size of matrix $A$ grows, the size of the data sets which define the equations increase rapidly over time. For classical algorithms such as the Conjugate Gradient Method~\cite{shewchuk94}, it takes about total runtime of $O(Ns\sqrt \kappa  \log (1/\varepsilon ))$ to get the solution $\vec x$ when $A$ is positive definite or $O(Ns\kappa \log (1/\varepsilon ))$ when $A$ is not. For all classical algorithms, it is shown that the linear lower bound $O(N)$ in runtime scaling cannot be broken even we are only interested in $\vec{x}^\dagger{M}\vec{x}$ rather than $\vec{x}$ itself~\cite{originalpaper}.

The quantum algorithm proposed by Harrow, Hassidim and Lloyd \cite{originalpaper} for solving linear systems of equations is shown to improve the runtime scaling to the $O(\log{N})$ regime. The algorithm starts with a quantum state $|b\rangle$ and a few ancilla qubits in state $|0\rangle$. Here the state vector of $|b\rangle$ in the computational basis represents the unit vector $\vec{b}$. Applying the well-known phase estimation subroutine~\cite{nielsen00} on $|0\cdots{0}\rangle|b\rangle$ with $U=e^{-iAt_0}$ as the unitary operation, we obtain the final state of the two-register system which is approximately $\sum_j\beta_j|\lambda_j\rangle|u_j\rangle$ up to a normalization constant ({When the eigenvalues of $A$ can be exactly encoded using the ancilla bits, which is the case of our experiment as we will see later in the paper, the final state of the phase estimation is exactly proportional to $\sum_j\beta_j|\lambda_j\rangle|u_j\rangle$}). Here $|u_j\rangle$ is the eigenbasis of $A$ and $|b\rangle=\sum_j\beta_j|u_j\rangle$. When the phase estimation is completed, an ancilla bit is added to the system and a controlled rotation is performed on it with the $|\lambda_j\rangle$ register as the control register. The state of the system then becomes $\sum_j\left(\sqrt{1-{\tilde{C}}^2/\lambda_j^2}|0\rangle+{\tilde{C}}/\lambda_j|1\rangle\right)\beta_j|\lambda_j\rangle|u_j\rangle$ where ${\tilde{C}}$ is a normalization constant. By inverting all the previous quantum operations except for the controlled rotation on the ancilla bit, we uncompute the $|\lambda_j\rangle$ register back to the state $|0\cdots{0}\rangle$. Conditioning on measuring $|1\rangle$ in the ancilla qubit, The algorithm probabilistically outputs a state $|x\rangle=\sum_j\beta_j\lambda_j^{-1}|u_j\rangle$ in the register that is initially in the state $|b\rangle$. The state vector of $|x\rangle$ in the computational basis is proportional to the solution $\vec{x}$ of the linear system $A\vec{x}=\vec{b}$. The total number of quantum operations needed for the algorithm is $O({s^2}{\kappa ^2}\log N)$. This implies an exponential speedup over the Conjugate Gradient Method,  which is the best classical algorithm for solving the problem.

Let us consider a $2\times{2}$ system to find $\vec x=(x_1\quad x_2)^T$ such that $A\vec x  = \vec b$ where

\begin{equation}\label{eq:Axb}
A=\frac{1}{2}\begin{pmatrix}
3 & 1 \\
1 & 3 \\
\end{pmatrix};\qquad
\vec{b}=\begin{pmatrix}
{b_1} \\ {b_2} \\
\end{pmatrix}
\end{equation}.

where $\vec{b}$ is normalized. The spectral decomposition $A=\sum_{j=1}^2\lambda_j|u_j\rangle\langle{u_j}|$ has $\lambda_1=1$, $|u_1\rangle=\frac{1}{\sqrt{2}}(|0\rangle-|1\rangle)$ and $\lambda_2=2$, $|u_2\rangle=\frac{1}{\sqrt{2}}(|0\rangle+|1\rangle)$. The quantum algorithm to solve this problem can be summarized as the following six steps \cite{originalpaper} (see Fig. ~\ref{fig:demon}).

\begin{enumerate}

\item Input the vector $\vec b$ as a quantum state $\left| b \right\rangle  = \sum\limits_{i = 1}^N {{b_i}} \left| i \right\rangle$ stored in a quantum register (termed register $B$) and prepare the register $C$ in the state $\frac{1}{T}\sum\limits_{\tau  = 0}^{T - 1} {\left| \tau  \right\rangle }=\frac{1}{4}(|00\rangle+|01\rangle+|10\rangle+|11\rangle)$. Here $\left| i \right\rangle$ and $|\tau\rangle$ represents the computational bases of registers $B$ and $C$ that encode the value of $i$ with the $n=\log{N}=1$ qubit and $t=\log{T}=2$ qubits respectively.

\item Apply the conditional Hamiltonian evolution $\sum\limits_{\tau  = 0}^{T - 1} {\left| \tau  \right\rangle } {\left\langle \tau  \right|^c} \otimes {e^{-iA\tau {t_0}/T}}$ on both registers $B$ and $C$ up to error ${\varepsilon _H}$. Here `$c$' in the notation $\left| \tau  \right\rangle {\left\langle \tau  \right|^c}$ marks the control register and $t_0=2\pi$.

\item \label{step:pea} Apply quantum Fourier transform to the register $C$. Denote the Fourier basis as $\left| k \right\rangle$. In general, at this stage in the superposition state of both registers, the amplitudes of the basis states are concentrated on $k$ values that approximately satisfy ${\lambda _k} \approx \frac{{2\pi k}}{{{t_0}}}$, where ${\lambda _k}$ is the $k$-th eigenvalue of the matrix $A$. However, $A$ is defined as in Eq.~\eqref{eq:Axb}, the eigenvalues $\lambda_1$, $\lambda_2$ are exactly powers of 2. Therefore if we further let $t_0=2\pi$, $\left| k \right\rangle = \left| {{\lambda _k}} \right\rangle$, the phase estimation subroutine will produce exactly $\sum_j\beta_j|\lambda_j\rangle|u_j\rangle/(\sum_j\beta^2_j)^{1/2}$.

\item \label{step:inv} Add an ancilla qubit and apply conditional $y$-rotation $R_y$ on it, controlled by the register $C$. This rotation transforms the qubit to $\sqrt {1 - \frac{{{{\tilde{C}}^2}}}{{\lambda _{_j}^2}}} \left| 0 \right\rangle  + \frac{{\tilde{C}}}{{{\lambda _j}}}\left| 1 \right\rangle$, where ${\tilde{C}}$ is a nonzero normalization constant . This is a key step of the algorithm and it involves finding the reciprocal of the eigenvalue ${\lambda _j}$ quantum mechanically and $y$-rotate the qubit with an angle $\theta_j$ such that $\sin\theta_j=\frac{{\tilde{C}}}{\lambda_j}$. To achieve both, we first use a {\sf SWAP} gate (Fig. ~\ref{fig:demon}) to accomplish the transformation

\begin{equation}\label{eq:lam_inv}
\begin{array}{c}
|\lambda_1\rangle=|01\rangle\overset{\sf\tiny SWAP}{\longrightarrow}|10\rangle=|2\lambda_1^{-1}\rangle \\[0.1in]
|\lambda_2\rangle=|10\rangle\overset{\sf\tiny SWAP}{\longrightarrow}|01\rangle|=|2\lambda_2^{-1}\rangle
\end{array}
\end{equation}
\\*
in register $C$. Then we use the register $C$ with $|2\lambda_j^{-1}\rangle$ states as the control register to apply the controlled $R_y(\lambda_j^{-1})$ rotation on the ancilla qubit. As shown in Fig. ~\ref{fig:demon}, the controlled-$R_y(\theta)$ gates applies rotation with $\theta_j=(2\pi/2^{r})\lambda^{-1}_j$ where $r$ is a parameter. In this experiment we let $r=2$ and use the approximation $\\sin(\theta_j/2)\approx \theta_j/2$. With this approximation, after the controlled-$R_y$ rotation gates, the state of the system is close to
\begin{equation}\label{eq:finalstate}
\sum_j\beta_j\left(\sqrt{1-\frac{\tilde{C}^2}{\lambda_j^2}}|0\rangle+\frac{\tilde{C}}{\lambda_j}|1\rangle\right)|\lambda_j\rangle|u_j\rangle
\end{equation} and $\tilde{C}$ is 0.736 according to the calculation based on Ref.\cite{jmp}.

\item Uncompute register $B$ and $C$ by inverting the operation in the steps 1-3.

\item Measure the ancilla bit. If it returns 1, the register $B$ of the system is in the state $\sum\limits_{j = 0}^N {{\beta _j}{\lambda _j^{ - 1}}\left| {{u_j}} \right\rangle }$ up to a normalization factor, which is equal to the solution $\vec{x}$ of the linear system $A\vec{x} = \vec{b}$.

\end{enumerate}

\begin{figure}
\centering
\includegraphics[scale=0.2]{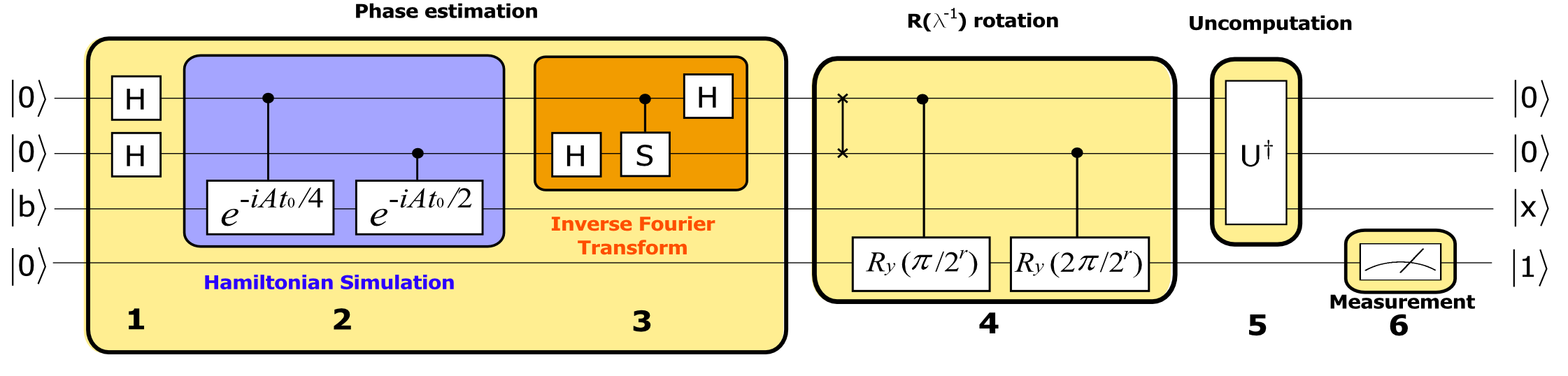}
\caption{(Color online) The four-qubit quantum circuit for the algorithm. The numeric labels {\sf 1} to {\sf 6} represent the 6 steps outlined in the text. Here $r=2$ and $t_0=2\pi$, determining the precision and probability of the correct answer. The matrix forms of the single-qubit gates are $S = \left( {\begin{array}{*{20}{c}}
   1 & 0  \\
   0 & i  \\
\end{array}} \right),\displaystyle H = \frac{1}{{\sqrt 2 }}\left( {\begin{array}{*{20}{c}}
   1 & 1  \\
   1 & { - 1}  \\
\end{array}} \right), \displaystyle R_y(\theta)=\left( {\begin{array}{*{20}{c}}
   \cos(\theta/2) & -\sin(\theta/2)  \\
   \sin(\theta/2) & { \cos(\theta/2) }  \\
\end{array}} \right)$.
 }
\label{fig:demon}.
\end{figure}

\begin{figure}
\includegraphics[scale=0.2]{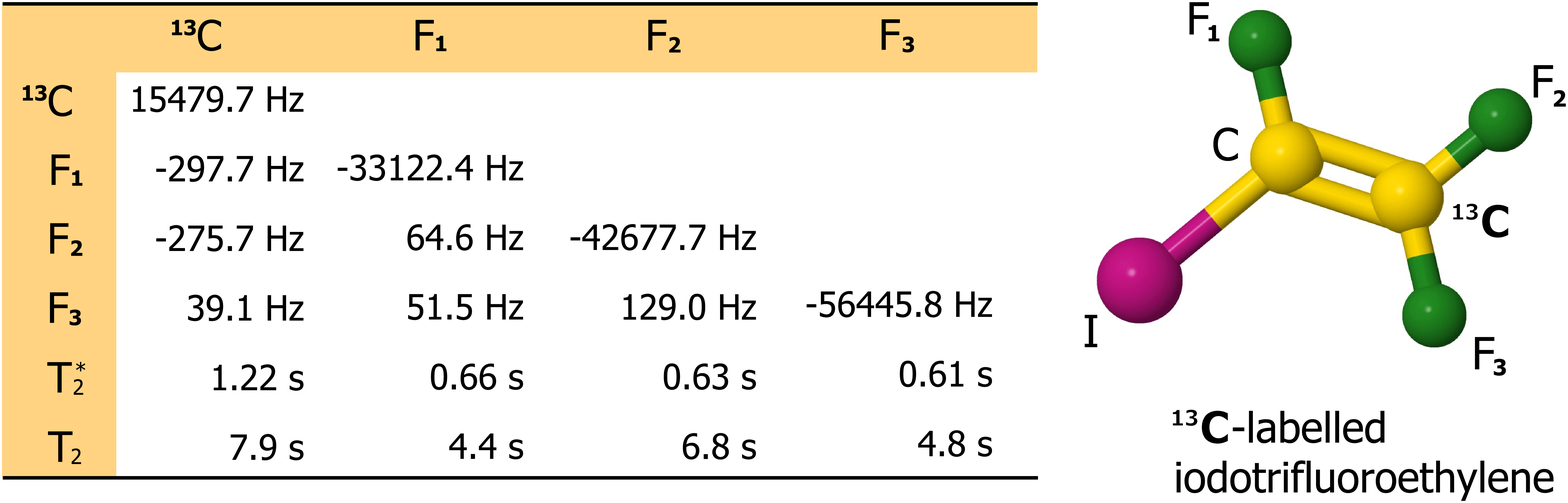}
\caption{(Color online) Properties of the molecular iodotrifiuoroethylene. The chemical shifts and J-coupling constants (in Hz) are on and below the diagonal in the table, respectively. The chemical shifts are given with respect to reference frequencies of 376.47 MHz (fluorines) and 100.64 MHz (carbons) at 303.0K. The transmitter offsets of carbon channel and fluorine channel are set at 15479.7Hz and -44701.0Hz respectively. The molecule contains four weakly coupled spin half nuclei which are ${}^{13}C,{}^{19}F_1,{}^{19}F_2,{}^{19}F_3$. The natural abundance of the sample with a single ${}^{13}C$ is about $1\%$.
 }
\label{fig:molecule}
\end{figure}

%\section{Experiment}
The experiment was carried out on a Bruker AV-400 spectrometer ($9.4 T$) at 303.0K. We chose iodotrifiuoroethylene dissolved in $d$-chloroform, where a $^{13}C$ nucleus and three $^{19}F _{1}$ nuclei consist of a four-qubit quantum system. We label $^{13}C$ as the first qubit, $^{19}F_{1}$, $^{19}F_{2}$ and $^{19}F_{3}$ as the second, third and forth qubit. Fig. 2 shows the measured properties of this four-qubit quantum system \cite{Luozhihuang}. We control the fluctuation of the temperature to be within 0.1K such that the effect of the chemical shift variations is suppressed. This system is first prepared into a pseudo-pure state (PPS) ${\rho _0}{\rm{ = }}\frac{{1{\rm{ - }}\varepsilon }}{{16}}I + \varepsilon \left| {0000} \right\rangle \left\langle {0000} \right|$ with $I$ representing the $16 \times 16$ unity operator and $\varepsilon  \approx {10^{{\rm{ - }}5}}$ the polarization, using the line-selective-transition method \cite{PPS, Luozhihuang}. It needs two gradient ascent pulse engineering (GRAPE) pulses \cite{grape} and two gradient field pulses. We apply quantum state tomography \cite{tomography} to reconstruct its experimental density matrix of the prepared PPS and the experimental fidelity is around 98.7$\%$.  The state fidelity is calculated by $F = {\rm Tr}\left( {{\rho _{\rm theory}}{\rho _{\rm exp}}} \right){\rm{/}}\sqrt {{\rm Tr}\left( {\rho _{\rm theory}^2} \right){\rm Tr}\left( {\rho _{\rm exp}^2} \right)}$, where ${{\rho _{\rm exp}}}$ and ${{\rho _{\rm theory}}}$  represent experimentally measured density matrices and the theoretical expectation, respectively. Then, we perform a rotation operation ${R_y}(\theta) = e^{-i I_y^4 \theta}$ (i.e.,a rotation along the $y$ axis with an angle $\theta$ to the $^{19}F_{3}$) \cite{quan_adia, quan_sim, cloning}, to obtain the initial state ${\rho _{\rm in}} = \left| {000b} \right\rangle \left\langle {000b} \right|$, where the normalized state $\vert b \rangle = \cos (\theta/2) \vert 0 \rangle + \sin (\theta/2) \vert 1 \rangle$ with the state vector $\vec{b} = [\cos (\theta/2) , \sin (\theta/2) ]^T$.

\begin{figure}
\centering

\includegraphics[width=0.8\textwidth]{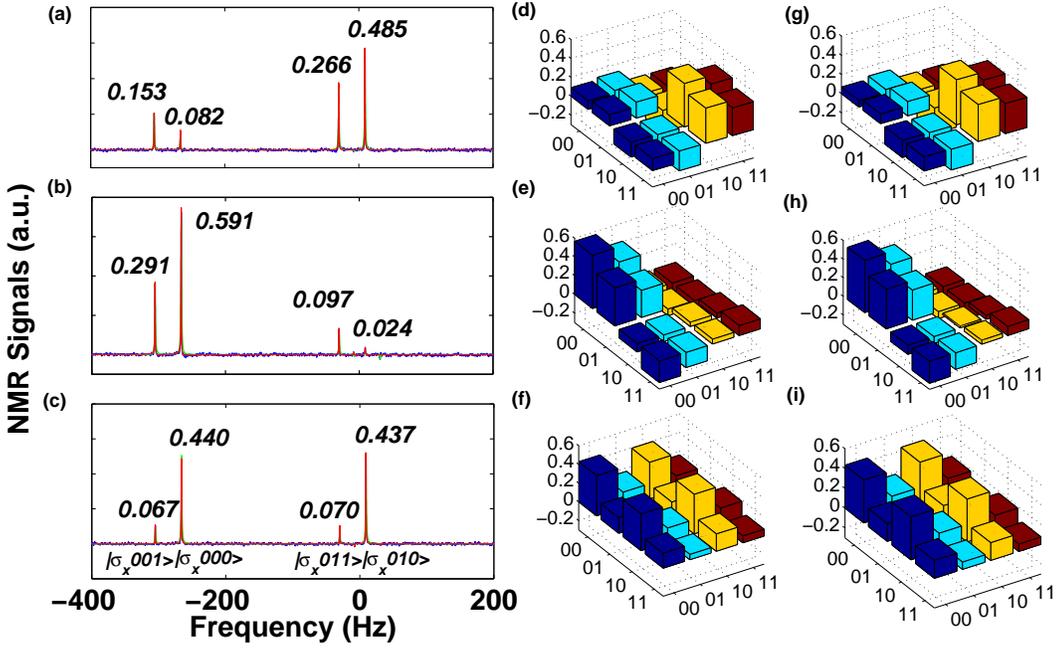}
\caption{(Color online) Experimental spectra and reconstructed density matrices for the final states. (a), (b), (c) Experimental $^{13}C$ spectra of the final states after a $\pi/2$ readout pulse for three different $\vec{b}$ which are listed in Fig.~\ref{fig:exp_res}.
There are eight peaks for carbon. Here we only show four of them related to the solution, and their intensities represent the respective probabilities of projecting the final state onto the states $\left| {0011} \right\rangle , \left| {0000} \right\rangle , \left| {0001} \right\rangle , \left| {0010} \right\rangle$ from left to right \cite{material}. The other four peaks are almost zero which are not shown here. The vertical axes have arbitrary but the same units. The numbers above the peaks are the relative intensity compared to the intensity of the peak of PPS. The ratio of the intensity of the peaks related to $\left| {0001} \right\rangle$ and $\left| {0011} \right\rangle$ approximates ${\left| x_1/x_2 \right|^2}$ of the solution $\vec x = (x_1\quad{x_2})^T$. We can obtain these ratios from (a), (b), (c) which are about 1:2, 3:1, 1:1 respectively. These agree with the theoretical values in Fig. ~\ref{fig:exp_res}. The experimentally measured, fitting and ideal spectra are shown as the blue, red and green curves, respectively. (d), (e), (f) Real parts of experimentally reconstructed density matrices of the final states in the subspace where the first and the second qubits are in the $\left| 0 0 \right\rangle$ state, along with the theoretical expectations (g), (h), (i). The rows and columns represent the standard computational basis in binary order, from $\left| {00} \right\rangle$ to $\left| {11} \right\rangle$. The intensities of the rest parts of the real parts and all the image parts are less than $3\%$ which can be seen in the supplementary material.}
\label{fig:nmr_res}
\end{figure}

\begin{figure}
\includegraphics[scale=0.6]{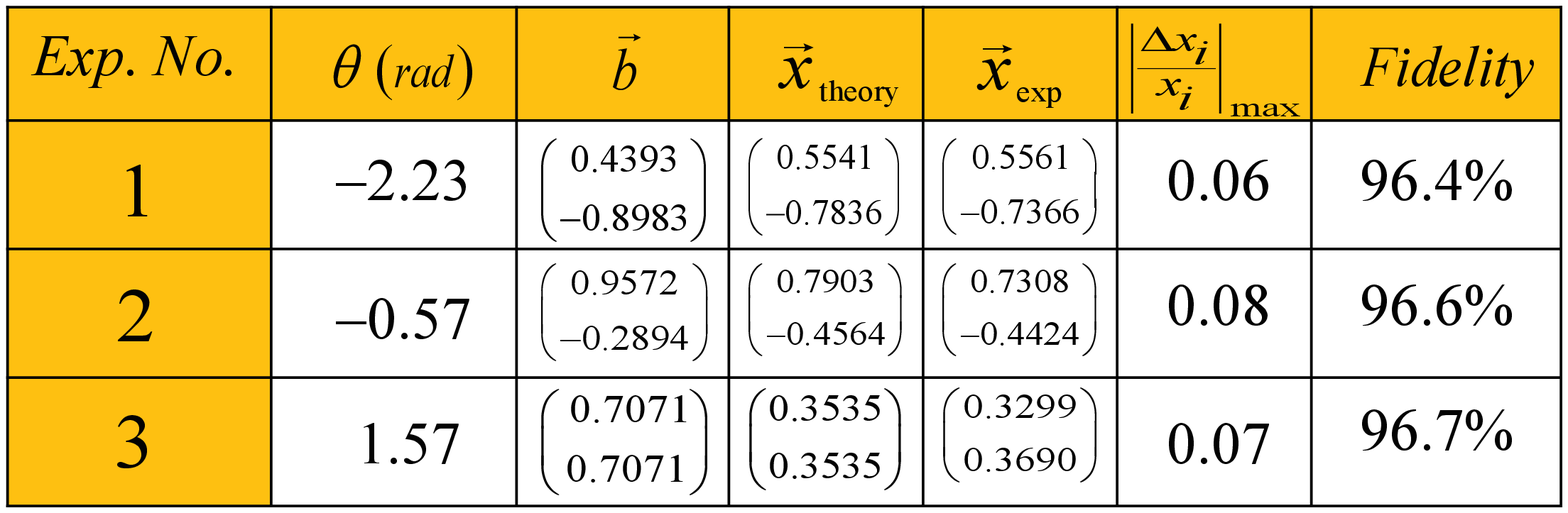}
\caption{(Color online)   Experimental results of quantum algorithm for solving the linear equations $A\vec x  = \vec b $. ${\vec{x}_{\rm theory}}$ is the theoretical solution while ${\vec x_{\exp }}$ is the experimental solution picked up from quantum state tomography. Here ${\left| {\frac{{\Delta {x_i}}}{{{x_i}}}} \right|_{max}} = {\left| {\frac{{x_{\exp }^i - x_{\rm theory}^i}}{{x_{\rm theory}^i}}} \right|_{max}}$, where ${x_{\exp }^i}$, ${x_{\rm theory}^i}$ are the $i$th elements of ${\vec  x_{\rm theory}}$ and ${\vec  x_{\exp }}$ respectively. The fidelity in the table refers to the final state of the whole 4-qubit system, and is measured between the experimental final state and the theoretical algorithm outcome.}
\label{fig:exp_res}
\end{figure}

Now, we implement the quantum circuit of the algorithm shown in Fig. \ref{fig:demon} on the prepared input state $\rho_{in}$. The quantum circuit is realized as a whole block using a shaped radio-frequency (r.f.) pulse that is optimized by the gradient ascent pulse engineering (GRAPE) algorithm \cite{grape,nmr_qc,factor143}. The GRAPE pulse is characterized by 1500 segments, the pulse duration of 22.5 ms, and is robust to r.f. inhomogeneities, with a theoretical fidelity 0.995. Finally, we measure the final state to obtain the solution of the linear equations. The desired final state is $\left| \Psi _{end}  \right\rangle  = {\rm{a}}\left| {0000} \right\rangle  + b\left| {0010} \right\rangle {\rm{ + }}\underbrace {{\rm{c}}\left| {0001} \right\rangle  + d\left| {0011} \right\rangle }_{\left| {00x1} \right\rangle }$ with $ \vert x \rangle = c\left| {0} \right\rangle + d\left| {1} \right\rangle$, the solution $\vec x =  (x_1 \quad{x_2})^T = (c \quad{d})^T/\tilde{C}$ is encoded into the state of qubit 3 in the subspace labeled by $\vert 0 \rangle_1 \vert 0 \rangle_2 \vert 1 \rangle_4$. Here we consider three different $\vec b$ by preparing three input states $\rho_{in}$ with different $\theta$. We perform a partial state tomography to get the information of $c$ and $d$ (see Method). The experimental $^{13}$C spectra are shown in the left side of Fig.  \ref{fig:exp_res}. $\left| x_1/x_2 \right|^2 = \left| c/d \right|^2 $ are the ratio of the probabilities of projecting the final state to the states $\left| {0001} \right\rangle$ and $\left| {0011} \right\rangle$, and the relative phase between ${{x_1}}$ and ${{x_2}}$ can be obtained by the coherence term $cd^*$ or $c^*d$ of qubit 3.
Since A and $\vec{b}$ both are real, $\vec{x}$ is real and the relative phase is either 0 or $\pi$ by the sign of $cd^*$ or $c^*d$. There may exist a global phase which can not be determined in the solution ${\vec x}$, but this global phase is often not important when one estimates the expectation value of some operator associated with $\vec{x}$. We list all results of the three experiments in Fig. \ref{fig:exp_res}.  The experimental errors are about $7\%$ which is measured by the formula ${\left| \Delta{x_i}/x_i \right|_{\rm max}}$. Furthermore, we perform the complete quantum state tomography \cite{tomography} for the final state in the Hilbert space spanned by these four qubits which is described in supplementary materials. The real parts of the reconstructed density matrices in the subspace labelled by $\vert 00 \rangle_{12}$ are shown in the right side of Fig. \ref{fig:nmr_res}, with the experimental fidelities are all above $96\%$.

The errors of the experiment mainly come from the inhomogeneity of magnetic fields, the imperfection of the GRAPE pulses and the chemical shift variations. The imperfection of the GRAPE pulses produces about $1\%$ error to the final state which can cause about $2\%$ error to the intensities of some peaks. The experimental time is about 50 ms, about $10\%$ of the coherence time (${T_2}^*$ in Fig. 2). Hence, the decay of the signals due to the limitation of coherence time is an important source of errors. The chemical shift of fluorine changes by about $3$ Hz when the temperature varies by 1K \cite{material}, so we control the fluctuation of the temperature below 0.1K to reduce the error to about $0.2\%$. We obtain the intensity of the peaks by fitting the spectra using the Lorentzian curves to reduce the error by the noise. The total derivation between the experimental final state and the theoretical algorithm outcome is about 3$\%$.

The theoretical errors of algorithm concentrate on the phase estimation in the second step and the control-rotation operation in the forth step. In our cases, for the reason that the eigenvalues of $A$ which we chose satisfy ${\lambda _k} = \frac{{2\pi k}}{{{t_0}}}$, the phase estimation is perfect and produces no error in theory \cite{originalpaper}. Most theoretical errors come from the control-rotation operation which depends on the positive integer parameter $r$. This error essentially depends on the quality of the approximation $\sin \alpha \approx \alpha$, where $\alpha$ is the rotation angle which is integral multiple of $\frac{\pi }{{{2^{r + 1}}}}$ acted on the ancillary bit. The theoretical error in this step decreases as $r$ increases when the intensities of the counterpart peaks of the final states are inversely proportional to ${r^2}$ \cite{jmp}.  Therefore, we take a balanced choice $r=2$ in the experiments, which causes that the theoretical error is $4\%$ contributing to ${\left| {\frac{{\Delta {x_i}}}{{{x_i}}}} \right|_{\max }}$ and the intensities of the counterpart peaks are about $10\%$ of the intensity of PPS. As presented in Fig. \ref{fig:exp_res}, we know that the theoretical error caused by this approximation is one of the major errors.

%When $r=2$, the intensities of the counterpart peaks are about $10\%$ of the intensity of PPS. If $r=3$ or bigger, the intensities of the counterpart peaks will be less than $3\%$, which is hard to measure accurately in experiment since the amplitude of the peaks we wanted have the same scale as or less than the fluctuation of the peaks caused  by the experimental error. When $r=1$, although the intensity of the peaks is stronger, the theoretical error will be about $50\%$.

%\section{Conclusion}
In conclusion, we experimentally demonstrate for the first time the quantum algorithm for solving linear systems of equations in a 4-qubit NMR system. We acquire the solutions with errors about $7\%$ for all the three group experiments, which indicates fine experimental accuracy and the validity of the algorithm. This quantum experiment can be a good evidence that the quantum computer can help solving common and important problems. For solving linear equation plays important role in many classical algorithm and classical computer, this experiment is the first step to create and realize more quantum algorithm. It gives us the hope to broaden the application of quantum computer.

The application of the quantum linear equation solver could be extended to a range of applications in many fields. For example, the ability of using the quantum algorithm to solve Poisson equation~\cite{poisson} would allow quantum chemists to speed up electrostatic calculation in density function theory. Furthermore, since the quantum algorithm could be used for efficiently solving linear systems of differential equations \cite{berry}, quantum computer might prove useful for solving the differential equations systems that arise in technical application.

\textbf{Methods}

\smallskip{}

\textbf{Hamiltonian simulation.} In this experiment, since the unitary evolution $\exp(-iAt)$ is a single-qubit gate, when realizing the $\exp(-iAt)$ operations we are \emph{not} directly implementing any Hamiltonian simulation algorithms in the literature (such as \cite{BACS07}) which are responsible for the exponential speedup of the quantum algorithm~\cite{originalpaper}. Instead, we decompose the $\exp(-iAt)$ gates into elementary quantum gates using Group Leader optimization algorithm~\cite{groupleader}.

\textbf{Finding the reciprocal of the eigenvalue ${\lambda _j}$.} The inversion technique which we use in Eq.~\eqref{eq:lam_inv} to find the reciprocals of $\lambda_j$ is applicable only to some particular type of matrices $A$ such as the one defined in Eq.~\eqref{eq:Axb}, where a {\sf SWAP} gate combined with a proper choice of $r$ can give us a sufficiently good approximation of the amplitude $\tilde{C}/\lambda_j$ in the final state Eq.~\eqref{eq:finalstate} of the ancilla qubit. For the general case where the $t$-qubit register $C$ holds a superposition of states that are approximately $|\lambda_j\rangle$, one can resort to techniques such as Newton iteration for finding the reciprocals of $\lambda_j$. The quantum algorithm implementing Newton iteration has also been proven efficient~\cite{poisson}. Furthermore, in step~\ref{step:inv} we also used the approximation $\theta_j\approx\sin\theta_j$. In general this approximation does not hold but one could use bisection method \cite{poisson} to efficiently find $\theta_j=\arcsin(\tilde{C}/\lambda_j)$.

\textbf{Partial Tomography.} The natural abundance of the sample in which just one carbon is ${}^{13}C$ is about $1\%$. To distinguish those molecules against the large background, we read out all three qubits via the ${}^{13}C$ channel, by applying SWAP gates and reading out the ${}^{13}C$ qubit. The solution $\vec x$ can be obtained from the subspace labeled by $\vert 0 \rangle_1 \vert 0 \rangle_2 \vert 1 \rangle_4$, i.e., the information of $c$ and $d$. This can be achieved by 5 readout pulses ($YEEE$,$YEEE*SWAP_{12}$, $YEEE*SWAP_{13}$, $YEEE*SWAP_{14}$, $XEEE*SWAP_{13}$). Here $E$ represents the unity operator. $X$ and $Y$ represent, respectively, a ${\pi/2}$ rotation operation along $x$ and $y$ axis. $SWAP_{ij}$ denotes a SWAP operation between $i$th and $j$th qubits \cite{nielsen00}. The first four readout pulses are to get the probabilities of $|c|^2$ and $|d|^2$, while the last readout pulse is to get the relative phase of $c^* d $.

\begin{acknowledgments}
The authors would like to thank the support of National Key Basic Research Program of China (Grant No. 2013CB921800) and the Strategic Priority Research Program (B) of the Chinese Academy of Sciences (Grant No. XDB01030400)¡£ This work is also supported by the National Natural Science Foundation of China (Grant no. 91021005, 21073171, 11275183, 11104262, 10975124) and the Ministry of Education of China (Grant no. 20113402110044 and the Scientific Research Foundation for the Returned Overseas Chinese Scholars). Prof. Sabre Kais thanks the NSF Center for Quantum Information and Computation for Chemistry, Award number CHE-1037992.
\end{acknowledgments}

\section*{author contributions}
J. Pan, X. Yao, Z. Li, C. Ju, X. Peng and J. Du designed and implemented the experiment for realizing the quantum circuit. Y. Cao and S. Kais devised the quantum circuit based on the original paper by Harrow et al. All authors discussed the results and co-wrote the manuscript.

\section*{Competing financial interests}
The authors declare no competing financial interests.

\newpage
\section*{Supplementary Materials}

\section*{Appendix A. Pulse sequence of the experiment}

Figure 5 is the pulse sequence of the experiment. The first two GRAPE pulses and two gradient field pulses were used to prepare a pseudo-pure state (PPS) $\left| {0000} \right\rangle \left\langle {0000} \right|$. The third GRAPE pulse is to perform the quantum circuit. The time of the readout pulses differs from 0.4ms to 25ms. The implementation time of circuit and the readout pulse is about 0.05ms in total.

\begin{figure}[!htb]
\centering
\includegraphics[width=0.8\textwidth]{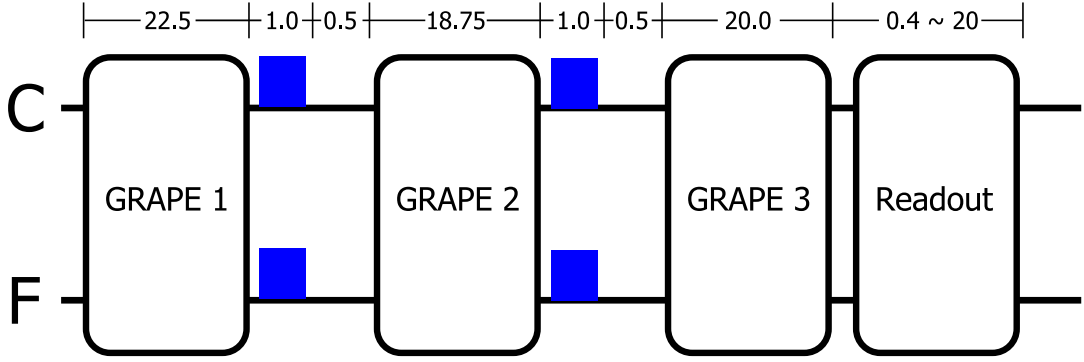}
\caption{\label{fig:demon} Pulse sequence of the experiment (not to scale). The numbers that label the pulse symbols represent the time duration of the pulse in milliseconds. The symbols `C' and `F' represent the channels for Carbon and Flourine atoms respectively. The rectangular boxes in the figure represent gradient field pulses.
 }

\end{figure}

\section*{Appendix B. Change of chemical shift caused by temperature fluctuation}

The chemical shifts are given with respect to reference frequencies
of 376.47 MHz (fluorines) and 100.64 MHz (carbons) at 303.0K. The chemical shifts of fluorines change by about 3 Hz when the temperature varies by 1K while the chemical shift of carbon is almost the same. The expression for the relationship between the chemical shifts of fluorines and the temperature is obtained by fitting the experimental data:
\[\begin{array}{l}
 {\omega _{{F_1}}} =  - 33122.4 - 3.0(T - 303.0) \\
 {\omega _{{F_2}}} =  - 42677.7 - 1.3(T - 303.0) \\
 {\omega _{{F_3}}} =  - 56445.8 + 1.6(T - 303.0) \\
 \end{array}\]
where ${\omega _{{F_1}}}$, ${\omega _{{F_2}}}$, ${\omega _{{F_3}}}$ represent the the chemical shifts of ${F_1}$, ${F_2}$, ${F_3}$ respectively and $T$ represents the temperature. The units are Hz and K. We controlled the temperature in $303.0 \pm 0.1K$ in the experiment.

\section*{Appendix C. State tomography.}
The state tomography for the whole 4-bit quantum state needs 44 readout pulses: $EEEE$, $EXEE$, $EYEE$, $EEXE$, $EXXE$, $EYXE$, $EEYE$, $EXYE$, $EYYE$, $EEEX$, $EXEX$, $EYEX$, $EEXX$, $EXXX$, $EYXX$, $EEYX$, $EXYX$, $EYYX$, $EEEY$, $EXEY$, $EYEY$, $EEXY$, $EXXY$, $EYXY$, $EEYY$, $EXYY$, $EYYY$, $swap_{12}*EEYY$, $swap_{12}*EEXY$, $swap_{12}*EEEY$, $swap_{12}*EEYX$, $swap_{12}*EEXX$, $swap_{12}*EEEX$, $swap_{12}*EEYE$, $swap_{12}*EEXE$, $swap_{12}*EEEE$, $swap_{13}*EEEY$, $swap_{13}*EEEX$, $swap_{13}*EEEE$, $swap_{14}*EEEE$, $YEEE$, $YEEE*swap_{12}$, $YEEE*swap_{13}$, $YEEE*swap_{14}$. ($E$ represents the unity operator. $X$ represents a rotation operation with the angle ${\pi/2}$ along $x$ axis while $Y$ along $y$ axis. $swap_{ij}$ means a swap operation between the $i$th and $j$th qubits.). As discussed in the methods, we can pick up the accurate and complete information of $\vec x$ using only 5 readout pulses: $YEEE$, $YEEE*swap_{12}$, $YEEE*swap_{13}$, $YEEE*swap_{14}$, $XEEE*swap_{13}$ to get the solution $\vec x$. Combining the first four readout pulses , we can obtain the diagonal elements of the density matrix. Using the fifth readout pulse, we can get ${{\rm{C}}_{24}}$ and ${{\rm{C}}_{42}}$. The state tomography results using 44 readout pulses for PPS and the experimental final states are showed in Figure 6 and Figure 7.

\begin{figure}

\centering
\includegraphics[width=0.8\textwidth]{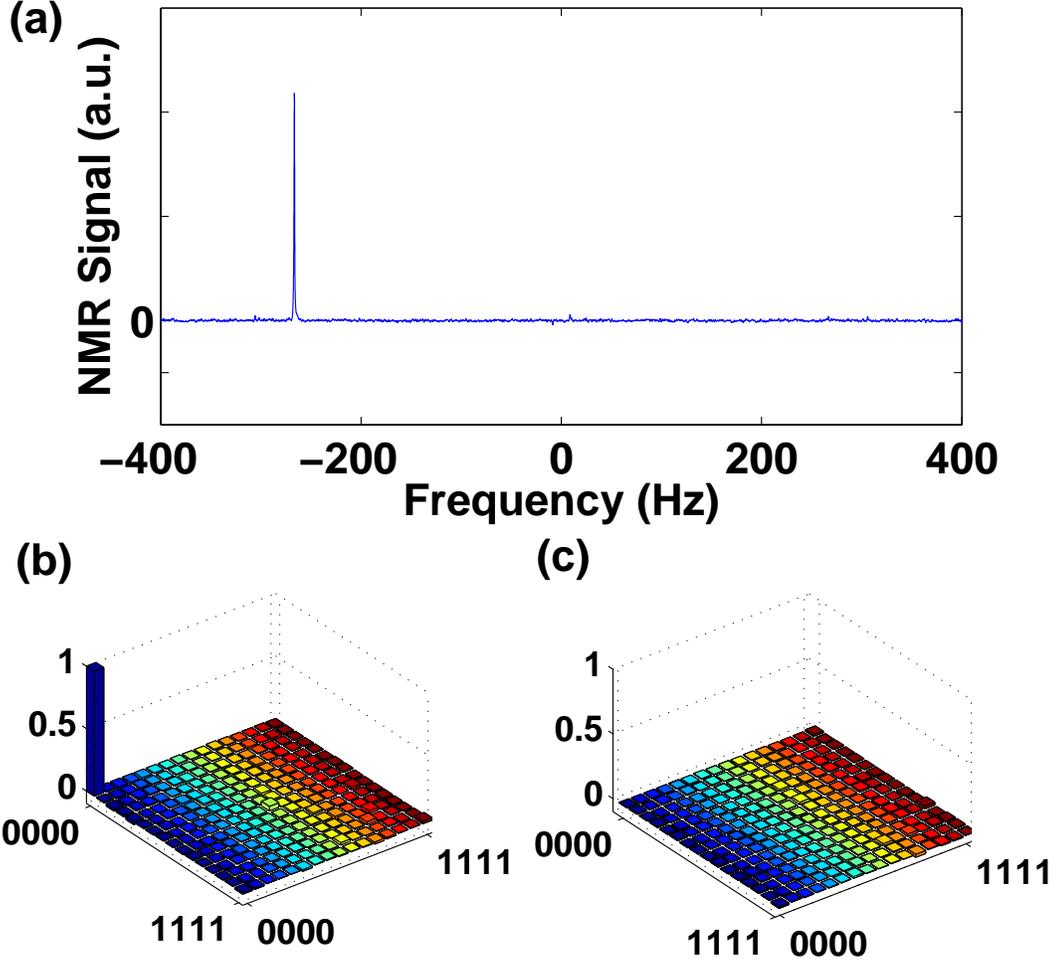}
\caption{\label{fig:demon}(Color online) Spectra and state tomography of the pseudo-pure state. (a) is the spectra of C obtained by $\pi/2$  readout pulse. The vertical axe have arbitrary unit. (b) is the real part of state tomography of PPS. (c) is the image part of tomography of PPS. The fidelity of the whole PPS is 98.73$\%$.
 }

\end{figure}

\begin{figure}

\centering
\includegraphics[width=0.8\textwidth]{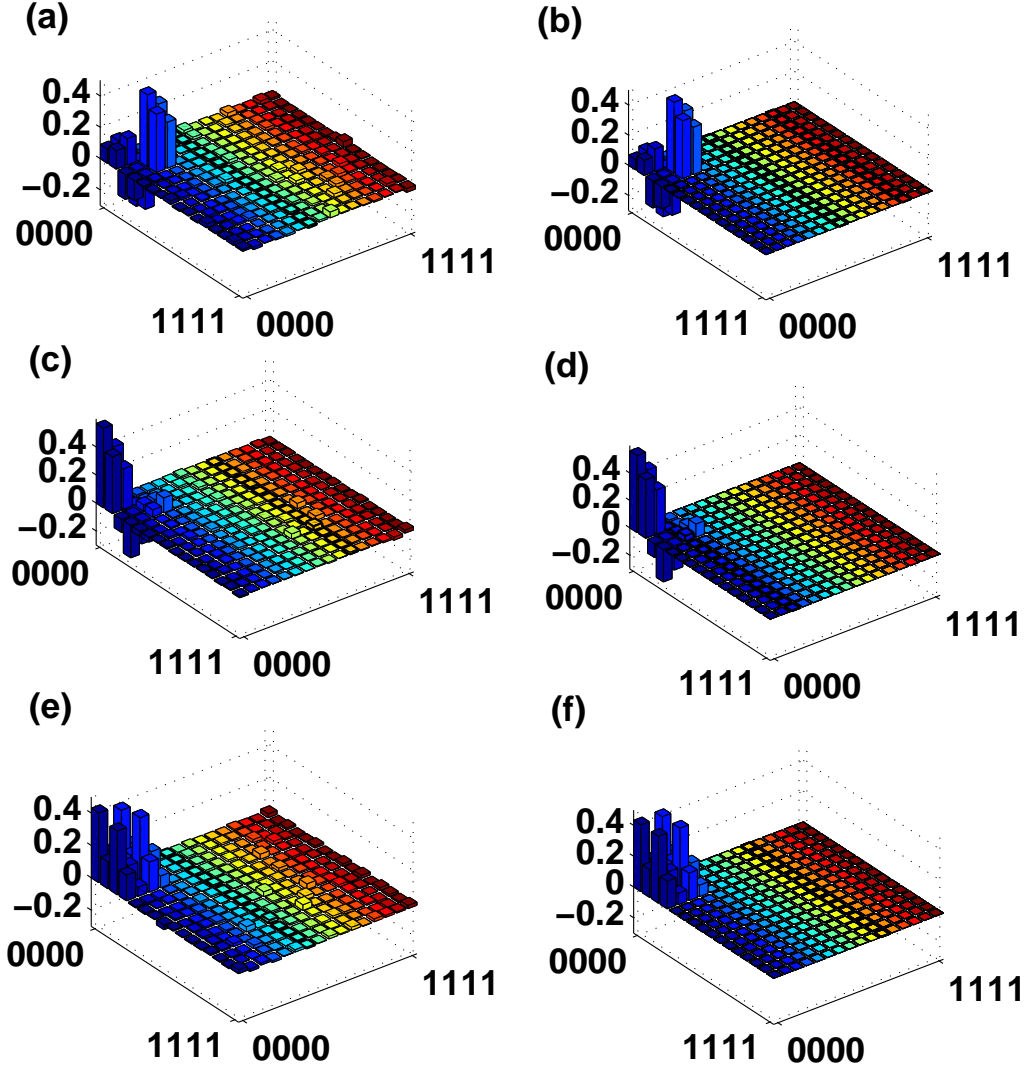}
\caption{\label{fig:demon}(Color online) Final state tomography. (a),(c),(e) are the real parts of the state tomography of the experimental final states for experiment 1,2,3 respectively, while (b),(d),(f) are the real parts of the state tomography of the theoretical final states. The rows and columns represent the standard computational basis in binary order, from $\left| {0000} \right\rangle$ to $\left| {1111} \right\rangle$. The fidelitis of the experimental final states are 96.4$\%$,96.6$\%$,96.7$\%$,respectively.
 }

\end{figure}

\section*{Appendix D. Relationship between peaks and elements of density matrix}
\begin{figure}
\centering
\includegraphics[width=0.8\textwidth]{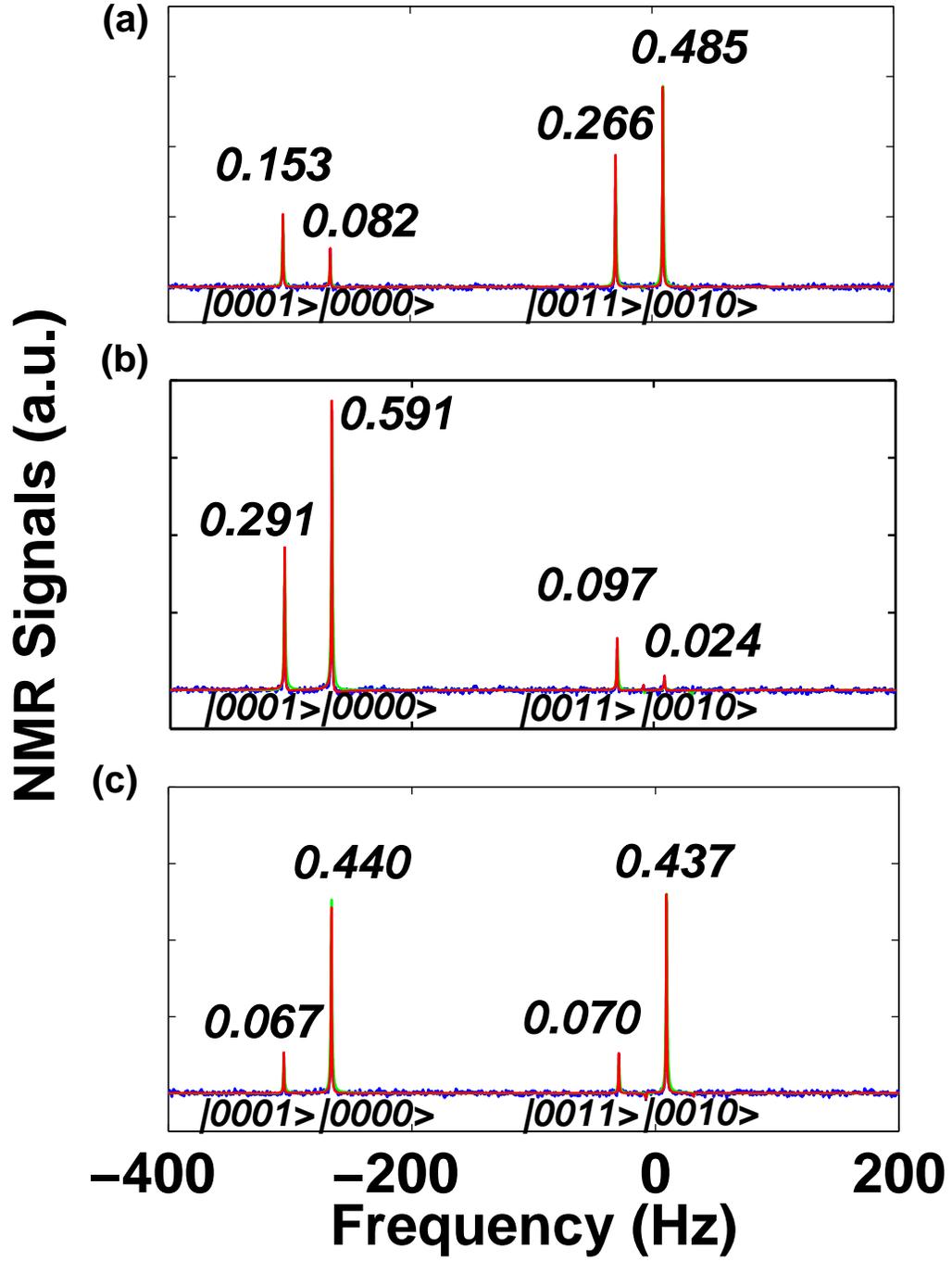}
\caption{\label{fig:demon}(Color online)Experimental spectra of C. (a),(b),(c) are the final state spectra by a gradient field pulse and a $\pi/2$  readout pulse in experiment 1,2,3 which are listed in table 2. The vertical axes have arbitrary but the same units.
 }

\end{figure}

As shown in Figure 8, there are eight peaks for carbon. Four of them are almost zero, and the intensity of the four significant peaks quantify the probabilities of the states $\left| {0001} \right\rangle$, $\left| {0000} \right\rangle$, $\left| {0011} \right\rangle$, $\left| {0010} \right\rangle$ from left to right respectively. In fact, defining the probabilities of the states $\left| {0000} \right\rangle$, $\left| {0001} \right\rangle$, ... , $\left| {1111} \right\rangle$ are ${p_0}$, ${p_1}$, ..., ${p_{15}}$, the four peaks which can be seen obviously are proportional to ${p_1} - {p_{9}}$, ${p_0} - {p_8}$, ${p_3} - {p_11}$, ${p_2} - {p_{10}}$ from left to right. Since ${p_i}\approx
0$ for all $i>4$ in experiment, the intensity of the four peaks are approximate proportional to the probabilities of the states $\left| {0011} \right\rangle$, $\left| {0000} \right\rangle$, $\left| {0001} \right\rangle$, $\left| {0010} \right\rangle$ respectively.

\end{spacing}
\end{document}